\documentclass[prl,superscriptaddress]{revtex4-2} 

\usepackage{physics} 
\usepackage{amssymb} 
\usepackage{mathptmx} 
\usepackage[cal=cm]{mathalfa} 
\usepackage{xcolor} 
\definecolor{deepblue}{rgb}{0, 0, 0.7} 
\usepackage[colorlinks]{hyperref}
\hypersetup{
    linkcolor=black,
    citecolor=deepblue, 
}

\newcommand{\up}[2]{^{\raisebox{#1pt}{$\sty#2$}}} 
\newcommand{\down}[2]{_{\raisebox{-#1pt}{$\sty#2$}}} 
\newcommand{\sty}{\scriptstyle}
\newcommand{\ssty}{\scriptscriptstyle} 
\newcommand{\move}[1]{\hspace{#1pt}}

\newcommand{\newcustomenv}[2]{
\newcounter{#1}
\newenvironment{#1}{
\refstepcounter{#1}
\par\emph{#2 \arabic{#1}:}
}{\par}
}

\newcustomenv{definition}{Definition}
\newcustomenv{theorem}{Theorem}
\newcustomenv{lemma}{Lemma}
\newcustomenv{corollary}{Corollary}

\usepackage{booktabs}
\usepackage{subcaption} 
\usepackage{graphicx}
\usepackage{forest}
\usepackage{tikz-qtree}

\begin{document}
    \title{Genuinely entangled subspaces beyond strongly nonlocal unextendible biseparable bases}

\author{Subrata Bera}
\email{98subratabera@gmail.com}
\affiliation{Department of Applied Mathematics, University of Calcutta, 92, A.P.C. Road, Kolkata- 700009, India}

\author{Atanu Bhunia}
\email{atanu.bhunia31@gmail.com}
\affiliation{Department of Mathematical Sciences, Indian Institute of Science Education and Research Berhampur, Laudigam, Konisi, Berhampur 760 003, Odisha, India}

\author{Indranil Biswas}
\email{indranilbiswas74@gmail.com}
\affiliation{Department of Applied Mathematics, University of Calcutta, 92, A.P.C. Road, Kolkata- 700009, India}

\author{Indrani Chattopadhyay}
\email{icappmath@caluniv.ac.in}
\affiliation{Department of Applied Mathematics, University of Calcutta, 92, A.P.C. Road, Kolkata- 700009, India}

\author{Debasis Sarkar}
\email{dsarkar1x@gmail.com,dsappmath@caluniv.ac.in}
\affiliation{Department of Applied Mathematics, University of Calcutta, 92, A.P.C. Road, Kolkata- 700009, India}

\begin{abstract}
Quantum information theory reveals a clear distinction between local and nonlocal correlations through the entanglement across spatially separated subsystems. The orthogonal complement of an unextendible biseparable basis (UBB) consists entirely of genuine multipartite entangled states, representing the most robust form of such nonlocal correlations. In this letter, we provide a sufficient condition for any subspace to be genuinely entangled, enabling the systematic construction of high-dimensional genuinely entangled subspaces (GESs) from UBBs. Our construction yields the largest known GES ever obtained from a UBB. In fact, every state in this subspace is 1-distillable across every bipartition which is one of the crucial result we obtained. Furthermore, we prove that every UBB is indistinguishable under LOCC protocols, underscoring a distinct manifestation of quantum nonlocality. The UBBs we construct exhibit strong nonlocality in this scenario, making cryptographic protocols secure not only against LOCC-based attacks but also against coordinated group attacks. We introduce a no-go condition that certifies such an extreme form of nonlocality. All previously known UBBs satisfy this condition, which may lead to the misconception that strong nonlocality is an inherent property of every UBB. However, we construct a UBB that violates the no-go condition and exhibits locality across certain bipartitions, challenging conventional notions of unextendibility and nonlocality in multipartite quantum systems.
\end{abstract} 
	
    %\date{\today}
%    \pacs{}
    \maketitle
    
%    {Keywords: Entanglement, indistinguishability, multipartite system, strong nonlocality, LOCC.}

%\textbf{\textit{Introduction}}---

Quantum nonlocality is a cornerstone of quantum information theory, underscoring the fundamental distinctions between classical and quantum correlations \cite{Einstein1935}. A key challenge in this field is the task of local state discrimination, in which spatially separated parties attempt to identify an unknown quantum state from a known set. If the states can be perfectly identified under Local Operations and Classical Communications (LOCC), the set is deemed locally distinguishable; otherwise, it is locally indistinguishable \cite{BennettUPB1999,Popescu2001,Xin2008,Walgate2000,Virmani,Ghosh2001,Ghosh2002,Groisman,Walgate2002,Divincinzo,Horodecki2003,Fan2004,Ghosh2004,Nathanson2005,Niset2006,Ye2007,Fan2007,Runyo2007,somsubhro2009,Feng2009,Yu2012,Yang2013,Zhang2014,somsubhro2009(1),somsubhro2010,yu2014,somsubhro2014,somsubhro2016,Divincenzo2000,Smolin2001,Chitamber2014}. Such locally indistinguishable sets of orthogonal pure states exhibit a striking form of nonlocality \cite{NonlocalityIndistinguishability} that is qualitatively distinct from Bell-type nonlocality \cite{Bell1964,Bruner2013}. Beyond its theoretical significance, it has profound implications for cryptographic primitives such as data hiding \cite{Terhal2001,Werner2002,chattopadhyay2007,Winter2009} and secret sharing \cite{Markham2008}. Its core feature of resilience against LOCC-based attacks provides inherent security against unauthorized access to encoded information.

A remarkable example of local indistinguishability arises from the concept of unextendible bases, specifically, the Unextendible Product Basis (UPB) introduced by Bennett \textit{et al.} \cite{BennettUPB1999} in 1999. They demonstrated that even a set of orthogonal product states remains indistinguishable under LOCC, establishing the notion of ``nonlocality without entanglement" \cite{BennettPB1999,Zhang2015,Wang2015,Chen2015,Yang2015,Zhang2016,Xu2016(2),Zhang2016(1),Xu2016(1),Halder2019strong nonlocality, Halder2019peres set,Xzhang2017,Xu2017,Wang2017,Cohen2008,Zhang2019,somsubhro2018,zhang2018,Yuan2020,bhunia2020,bhunia2022,linchen2016,johnston2013}. This discovery not only reshaped our understanding of local discrimination but also led to the identification of entangled subspaces in which every state is necessarily entangled \cite{BennettUPB1999,Divincinzo}. The natural multipartite generalization of this concept is the Unextendible Biseparable Basis (UBB), a set of orthogonal biseparable states whose complementary subspace contains only genuinely entangled states. Such subspaces becomes even more intriguing when they produce distillable entanglement across all bipartitions \cite{Agrwal2019,Atanu2024}, a key resource for quantum communication.

The construction of a UBB closely resembles that of a UPB. Consequently, this gives rise to the intuitive belief that a UBB, like a UPB, should also be indistinguishable under LOCC. Surprisingly, this fundamental aspect has remained unexplored in the literature. To address this gap, we propose a logical construction inspired by Bennett's foundational work on product bases \cite{BennettUPB1999}. The following two lemmas provide the necessary theoretical foundation to rigorously establish this result.

\begin{lemma}
A UBB is not completable even in a locally extended Hilbert space.
\label{not_complementable}
\end{lemma}

Proof: If a set of states is completable in a locally extended Hilbert space, then its complementary space in the extended Hilbert space is biseparable. By local projections the states on the complimentary space in the extended Hilbert space can be projected onto the complementary space in the original Hilbert space. These states are also biseparable since local projections do not create genuine entanglement. But we have a contradiction since the state complementary to a UBB is genuinely entangled by definition.

We are now ready to show that uncompletability with local Hilbert space extensions is a sufficient condition for the indistinguishability of a set of orthogonal biseparable states by any sequence of local positive operator valued measurements (POVMs) even with the help of classical communications among the observers.

\begin{lemma}
Given a set $S$ of orthogonal biseparable states on $\mathcal{H}=\bigotimes_{\ssty i=1}^{\ssty m} \mathcal{H}_{i}$ with $\dim \mathcal{H}_{i}=d_{ i}, i=1,2,\ldots,m$. If $S$ is distinguishable under local projective operator-valued measurements and classical communications, then it is completable in $\mathcal{H}$. If $S$ is distinguishable under local positive operator valued measurements and classical communications, then it can be completed in some extended space $\mathcal{H}'=\bigotimes_{\ssty i=1}^{\ssty m}(\mathcal{H}_i \oplus \mathcal{H}'_i)$.
\label{complementable}
\end{lemma}

Proof: We show how a local projective operator-valued measurement (PVM) protocol leads directly to a way to complete the set $S$. At some stage of their protocol, the parties may have been able to eliminate members of the original set of states $S$, and they may have mapped the remaining set of orthogonal states into a new set of orthogonal states $S'$. Determining which member they have in this new set uniquely determines which state of $S$ they started with. 

At this stage, party $i_{\ssty 0}$ performs an $l$ outcome PVM which is given by a decomposition of the remaining Hilbert space $\mathcal{K}=\mathcal{K}_{\text {else }} \otimes \mathcal{K}_{i_{0}}$ into a set of $l$ orthogonal subspaces $\{ \mathcal{K}_{\text {else }} \otimes P_j \mathcal{K}_{i_{0}}\}_{\ssty j=1}^{\ssty l}$ where $\mathcal{K}_{\text {else }}=\otimes_{\ssty i \neq i_{0}} \mathcal{K}_i$.

Let us consider a state $\ket{\alpha}_{\move{-2.5}\ssty \mathcal{A}} \otimes\ket{\beta}_{\move{-1.5}\ssty \mathcal{B}}$ in $S'$ which is product in $\mathcal{A}|\mathcal{B}$ partition ($i_{\ssty 0} \in  \mathcal{B}$). If the state lies in one of these subspaces, it will be unchanged by the measurement.  If not, it will be projected onto one of the states $\{\ket{\alpha}\otimes P'_{j}\ket{\beta}\}_{\ssty j=1}^{\ssty l}$, where $P'_j=P_j\otimes \mathbb{I}_{{\ssty\mathcal{B}\setminus}{i_{0}}}$. Let $S''$ be this new projected set of states, containing both the unchanged states in $S'$ as well as the possible projections of the states in $S'$. If one of the subspaces does not contain a member of $S'$, it can be completed directly. For the other subspaces, let us assume that each of them can be completed individually with product states orthogonal to members of $S''$. In this way, we have completed the projected $S'$ on the full Hilbert space $\mathcal{K}$, as these orthogonal-subspace completions are orthogonal sets and they are a decomposition of $\mathcal{K}$. However, we have now completed the set $S''$ rather than the set $S'$. Fortunately, one can replace the projected states $\ket{\alpha} \otimes P'_1\ket{\beta},\ldots,\ket{\alpha} \otimes P'_l\ket{\beta}$ by the original state $|\alpha\rangle \otimes$ $|\beta\rangle$ and $l-1$ orthogonal states by making $l$ linear combinations of the projected states. They are orthogonal to all other states as each $\ket{\alpha} \otimes P'_j\ket{\beta}$ was orthogonal, and they can be made mutually orthogonal as they span an $l$-dimensional space on the $i_{\ssty 0}$ side. Thus at each round of measurement, a completion of the set of states $S'$ is achieved assuming a completion of the subspaces determined by the measurement.

The tree of nested subspaces will always lead to a subspace that contains only a single state of the set, as the measurement protocol was able to tell the states in $S$ apart exactly. But such a subspace containing only one state can easily be completed and thus, by induction, we have proved that the original set $S$ can be completed in $\mathcal{H}$.

Finally, we note that a POVM is simply a PVM in an extended Hilbert space (this is Neumark's theorem \cite{Quantum1993}). Thus any sequence of POVMs implementable locally with classical communications is a sequence of local PVMs in extended Hilbert spaces and the preceding argument applies, leading to a completion in $\mathcal{H}^{\prime}$.\hfill$\sty\blacksquare$\par

\begin{theorem} 
Members of a UBB are not perfectly distinguishable under LOCC.
\end{theorem} 

Proof: If the UBB were measurable by POVMs, it would be completable in some larger Hilbert space by Lemma \ref{complementable}. But this is in contradiction with Lemma \ref{not_complementable}.

Thus, the members of a UBB cannot be perfectly distinguished by LOCC, confirming their intrinsic nonlocal nature. However, in the multipartite setting, the concept of such nonlocality becomes considerably richer and more nuanced. It requires not merely the presence of nonlocal correlations, but the genuineness of such nonlocality. 

A set is said to be genuinely nonlocal if it remains indistinguishable even when multiple parties are allowed to perform joint measurements \cite{Rout2019,Li2021,Lu2024}. A more stringent manifestation of this phenomenon arises in the form of local irreducibility \cite{Halder2019StrongNonlocality}, wherein no state can be eliminated through orthogonality-preserving local measurements (OPLM) \cite{Halder2018,BandyopadhyayHalderActivation2021, Bera2024}. This leads to the most extreme form of nonlocality, known as strong nonlocality, which is characterized by the set remaining locally irreducible across all possible bipartitions. Such sets exhibit heightened robustness in cryptographic and data-protection protocols, as their indistinguishability not only resists LOCC-based attacks but also survives coordinated group attacks involving subsets of parties.

In this context, it becomes essential to investigate whether UBBs also exhibit such strong nonlocality. At first glance, one might confidently assume so. This assumption stems from the fact that UBBs, unlike UPBs, can incorporate partially local states (entangled states that are products in at least one bipartition) in their construction. As a result, it appears reasonable to conclude that UBBs exhibit even higher nonlocality than UPBs. Furthermore, existing studies seem to reinforce this notion, as all known UBBs demonstrate strong nonlocality \cite{Agrwal2019, Atanu2024}. However, this intuition is incorrect. We present here a counterexample: a UBB that explicitly violates this assumption. 

In what follows, we first establish a sufficient condition for unextendibility. Consider a set of bipartite pure states $\mathcal{S}=\{\ket{\psi\down{1}{i}}_{\move{-1.5}\ssty AB}\}_{\ssty i=0}^{\ssty n-1}$, where $n$ denotes the number of states. Alice ($ A$) and Bob ($B$)  are the associated parties with $d_{\ssty A}=\dim A$ and $d_{\ssty B}=\dim B$ respectively. Then for any basis $\{\ket{k}_{\move{-1.5}\ssty A}\}_{\ssty k=0}^{\ssty d_{A}-1}$ of $ A$ and $\{\ket{l}_{\move{-1.5}\ssty B}\}_{\ssty l=0}^{\ssty d_{B}-1}$  of $ B$, $\ket{\psi\down{1}{i}}$ can be expressed as: 
\begin{equation}
%\begin{array}{c}
\ket{\psi\down{1}{i}}_{\move{-1.5}\ssty AB}=\sum_{\ssty kl} a_{\ssty kl}^{\ssty (i)}\ket{kl}_{\move{-1.5}\ssty AB},\quad a_{\ssty kl}^{\ssty (i)}\in \mathbb{C}
%\end{array}
\label{general_form}
\end{equation}
\vspace{-10pt}

Let us define some $n\cross n$ matrices $\Lambda_{\ssty kp|lr}^{\ssty AB}$ for each $k,p(> k)\in [d_{\ssty  A}]$; $l,r(> l)\in [d_{\ssty B}]$; and $\Gamma_{\move{-2}\ssty st}$ for each $ s,t(> s)\in [n]$, which can be expressed as:
\begin{equation}
(\Lambda_{\ssty kp|lr}^{\ssty    AB})\down{1}{\ssty ij} =
\begin{vmatrix}
a_{\ssty kl}^{\ssty(i)} & a_{\ssty kr}^{\ssty(i)} \\[3pt]
a_{\ssty pl}^{\ssty(j)} & a_{\ssty pr}^{\ssty (j)}
\end{vmatrix},\quad
(\Gamma_{\move{-2}\ssty st})\down{1}{\ssty ij}=
\begin{vmatrix}
\delta_{\ssty si} & \delta_{\ssty ti} \\[2pt]
\delta_{\ssty sj}& \delta_{\ssty tj}
\end{vmatrix}
\label{product_forming_matrices}
%\label{symmetrization_matrices}
\end{equation}
where $i,j\in [n]$. Here, $\delta_{\ssty si}$ stands for the Kronecker delta function. Note that $[d]$ represents the set $\{0,1,\ldots,d-1\}$ and $(M)_{\ssty ij}$ refers to the $(i,j)$-th element of the matrix $M$. These notations are used extensively throughout this letter.

The matrices defined above are said to be product-forming matrices and symmetrization matrices respectively. There are precisely ${^{d_{\ssty A}}C_{2}} \cross {^{d_{\ssty B}}C_{2}}$ product-forming matrices and ${^{n}C_{2}}$ symmetrization matrices. Relying on their structural properties, we construct the next theorem, which will be crucial for proving the unextendability of a set of orthogonal product states.

\begin{theorem}
The subspace spanned by $\mathcal{S}$ is entangled, if 
the above matrices spans the space of all $n\cross n$ matrices over $\mathbb{C}$.
\label{sufficient_entangled}
\end{theorem}

Proof: An arbitrary state of the space spanned by $\mathcal{S}$ can be expressed as:
\vspace{-10pt}
\begin{equation*}
\ket{\chi}_{\move{-1.5}\ssty AB}=\sum_{ i=0}^{ n-1} x_{\ssty i}\ket{\psi\down{1}{i}}_{\move{-1.5}\ssty AB}=\sum_{ikl}x_{\ssty i} a_{\ssty kl}^{\ssty (i)}\ket{kl}_{\move{-1.5}\ssty AB}=\sum_{k=0}^{\ssty d_{\ssty A}-1}\ket{k}_{\move{-1.5}\ssty A}\ket{\phi\down{1}{k}}_{\move{-1.5}\ssty B},\quad x_{\ssty i}\in \mathbb{C}
\label{arbitrary_state}
\end{equation*}
$\vspace{-10pt}$

The state $\ket{\chi}$ is a product state if and only if the vectors $\ket{\phi\down{1}{k}}=\sum_{\ssty il}x_{\ssty i} a_{\ssty kl}^{\ssty (i)}\ket{l}\in\mathcal{H}^{\ssty B}$ (for $k\in [d_{\ssty A}]$) are pairwise linearly dependent. This is equivalent to collinearity between two vectors $(\sum_{\ssty i}x_{\ssty i}a_{\ssty kl}^{\ssty (i)})\ket{l}+(\sum_{\ssty i}x_{\ssty i}a_{\ssty kr}^{\ssty (i)})\ket{ r}$ and $(\sum_{\ssty i}x_{\ssty i}a_{\ssty pl}^{\ssty (i)})\ket{ l}+(\sum_{\ssty i}x_{\ssty i}a_{\ssty pr}^{\ssty (i)})\ket {r}$ for each $k,p(>k)\in[d_{\ssty  A}]$, and $l,r(> l)\in[d_{\ssty  B}]$. Therefore, 
\begin{equation}
\begin{vmatrix}
\underset{\ssty i}{\sum}\,x_{\ssty i} a_{\ssty kl}^{\ssty (i)} & \underset{\ssty i}{\sum}\,x_{\ssty i} a_{\ssty kr}^{\ssty (i)} \\[7pt]
\underset{\ssty i}{\sum}\,x_{\ssty i} a_{\ssty pl}^{\ssty (i)} & \underset{\ssty i}{\sum}\,x_{\ssty i} a_{\ssty pr}^{\ssty (i)}
\end{vmatrix}=0 
\label{2x2_block}
\end{equation}

Let us consider $x_{\ssty i}x_{\ssty j}=x_{\ssty ij}$, for $i,j \in[n]$. Then the Eqs. (\ref{2x2_block}) become linear in $x_{\ssty ij}$ with coefficient $(a_{\ssty kl}^{\ssty (i)}a_{\ssty pr}^{\ssty (j)}-
a_{\ssty kr}^{\ssty (i)}a_{\ssty pl}^{\ssty (j)})=(\Lambda_{\ssty kp|lr}^{\ssty   AB})\down{1}{\ssty ij} $:
\vspace{-10pt}
\begin{equation}
\sum_{\ssty ij}x_{\ssty ij}(\Lambda_{\ssty kp|lr}^{\ssty   AB})\down{1}{\ssty ij} =0
\label{product_forming_equation}
\end{equation}
\vspace{-10pt}

Also, by definition  $x_{\ssty st}=x_{\ssty ts}$, i.e., $x_{\ssty st}-x_{\ssty ts}=0$  for each $ s,t(>s)\in[n]$, which is equivalent to the linear equation 
\begin{equation}
\sum_{\ssty ij}x_{\ssty ij}(\Gamma_{\move{-2}\ssty st})\down{1}{\ssty ij} =0
\label{symmetrization_equation}
\end{equation}
\vspace{-10pt}

Associated to each nonzero solution of the Eqs. (\ref{2x2_block}) there exists a nonzero solution of Eqs.  (\ref{product_forming_equation}) and (\ref{symmetrization_equation}). On the other hand the system of homogeneous linear Eqs. (\ref{product_forming_equation}) and (\ref{symmetrization_equation}) do not have any nonzero solutions iff the matrices $\Lambda_{\ssty kp|lr}^{\ssty  AB}$ and $\Gamma_{\move{-2}\ssty st}$ spans the entire space of  $n\cross n$ matrices over $\mathbb{C}$, commonly denoted as $M(n,\mathbb{C})$. This proves the Theorem \ref{sufficient_entangled}.\hfill$\sty\blacksquare$\par

Now, imagine a multipartite system and consider a bipartition $\mathcal{P}$. Given a set of states $\mathcal{U}$, one can always determine the remaining orthogonal states and express them in the form of Eq. (\ref{general_form}). From this representation, it is possible to identify all associated product-forming and symmetrization matrices. If these matrices satisfy Theorem \ref{sufficient_entangled}, it can be concluded that $\mathcal{U}$ is unextendible in all the subpartitions of $\mathcal{P}$.
\begin{corollary} 
A set of pure biseparable states forms a UBB if its orthogonal complement satisfies Theorem \ref{sufficient_entangled} in every bipartition.
\label{sufficient_UBB}
\end{corollary}

This is the condition by which we will prove the unextendibility of a UBB throughout this letter. Interestingly, this condition is also sufficiently capable of proving that preexisting UBBs are unextendible.

Before proceeding further, we introduce some notions relevant to the construction of UBBs. Consider a pure state $\ket{\tau_{\ssty d}^{\move{0.7} r}}=(\sum_{\ssty i=0}^{\ssty d-1}\ket{i})^{\otimes r}$ in $d^{\otimes r}$ Hilbert space, commonly known as stopper state in the perspective of constructing some unextendible bases. Let $\mathcal{S}$ be a set of $n$ linearly independent pure states and $\ket{\psi}\in \mathcal{S}$ such that $\braket{\psi}{\tau_{\move{-0.3}\ssty d}^{\move{0.6} r}}\neq0$. Then, $\Omega_{\ssty d}^{\move{.5}r}(\mathcal{S},\ket{\psi})=\{(\move{2}\ketbra{\phi}{\psi}-\ketbra{\psi}{\phi}\move{2})\ket{\tau_{\ssty d}^{\move{0.7} r}}, \ket{\phi}\in \mathcal{S}\setminus\{\ket{\psi}\}\}$ is a set of $n-1$ linearly independent states perpendicular to $\ket{\tau_{\ssty d}^{\move{0.7} r}}$. In what follows, we will extensively use the states from this set to span the complementary subspaces associated with the forthcoming unextendible bases.

Now, let us define a set of orthogonal pure states $\mathcal{G}=\{\ket{\phi\down{1}{ i}}^{\move{-1}\ssty-},\ket{\psi\down{1}{ i}}^{\move{-1}\ssty\pm},\ket{\eta}^{\move{-1}\ssty\pm}\}_{  i=0}^{ 1}$, where 
\begin{equation}
\begin{array}{ll}
\ket{\eta}^{\move{-1}\ssty\pm} &\move{-3}= \ket{01\pm10}\ket{1},\ket{\phi\down{1}{ i}}^{\move{-1}\ssty\pm} = \ket{i}\ket{00\pm 10},\\
\ket{\psi\down{1}{ i}}^{\move{-1}\ssty\pm} &\move{-3}= \ket{i}\ket{i,1}+\omega_{\move{1}\ssty 2}^{\ssty\pm}\ket{\phi\down{1}{ i}}^{\move{-1}\ssty+},\text{with } \omega^{\ssty+}_{\ssty z}=1,\omega^{\ssty-}_{\ssty z}=-1/z.
\end{array}
\label{genuinely_local_basis}
\end{equation}

The set $\mathcal{G}$ forms a full basis in tripartite Hilbert space $\mathcal{H}^{\ssty 2\otimes 2 \otimes 2}_{\ssty ABC}$ unless we introduce $\ket{\tau_{\ssty 2}^{\ssty 3}}$, which neglects all the states of $\mathcal{G}\up{1}{\ssty +}=\{\ket{\psi\down{1}{ 0}}^{\move{-1}\ssty+},\ket{\psi\down{1}{ 1}}^{\move{-1}\ssty+},\ket{\eta}^{\move{-1}\ssty+}\}$ from $\mathcal{G}$. The new set $\mathcal{U}=(\mathcal{G}\setminus\mathcal{G}\up{1}{\ssty +})\hspace{-0.1em}\bigcup\{\ket{\tau_{\ssty 2}^{\hspace{1pt}\ssty3}}\}$ is generated in this way which contains $6$ mutually orthogonal biseparable states including the stopper state $\ket{\tau_{\ssty 2}^{\ssty 3}}$ itself. Accordingly, the states of $\Omega_{\ssty 2}^{\ssty 3}(\mathcal{G}\up{1}{\ssty +},\ket{\eta}^{\move{-1}\ssty+})$ spans the complementary space of $\mathcal{U}$. And they staisfy Theorem \ref{sufficient_entangled} in $A|BC$, $B|CA$ and $C|AB$ bipartitions. For each bipartition, the dimension of the space spanned by the product-forming matrices and symmetrization matrices turns out to be $4$, which is exactly the square of the number of states in $\Omega_{\ssty 2}^{\ssty 3}$. This implies these matrices spans $M(2,\mathbb{C})$. Therefore, by Corollary \ref{sufficient_UBB} the set $\mathcal{G}\up{1}{\ssty +}$ constitutes a UBB in the $2\otimes2\otimes2$ Hilbert space.

The following theorem establishes that not every UBB exhibits the highest form of nonlocality.

\begin{theorem} 
The set $\mathcal{U}$ is neither strongly nonlocal nor genuinely nonlocal.
\end{theorem}

Suppose Bob and Charlie go first with two outcome joint PVM $\{\ketbra{\tau_{\ssty 2}^{\ssty 2}}{\tau_{\ssty 2}^{\ssty 2}},\mathbb{I}-\ketbra{\tau_{\ssty 2}^{\hspace{1pt}\ssty 2}}{\tau_{\ssty 2}^{\ssty 2}}\}$. For every outcome, at least one state of $\mathcal{U}$ got eliminated. This proves $\mathcal{U}$ is reducible in $A|BC$ partition and consequently not strongly nonlocal.  The next part follows by observing the post-measurement states that are locally distinguishable in the $A|BC$ partition.

This result clarifies that the intuition assuming UBBs are necessarily strongly nonlocal is incorrect. Consequently, finding a strongly nonlocal UBB remains a nontrivial task. In the next part, we construct examples of strongly nonlocal UBBs that are significant not only for exhibiting the highest form of nonlocality but also because their complementary spaces are distillable across every bipartition. 

Consider the sets of orthogonal biseparable pure states:
\begin{equation*}
\begin{array}{ll}
\mathcal{G}\up{1}{\ssty +}_{\ssty hqm}=
&\bigl\{
\ket{\psi\down{1}{\ssty 1}}\up{-1}{\move{-1}\ssty (h)+},\ket{\eta\down{1}{\ssty 0}}\up{-1}{\move{-1}\ssty (q)+},\ket{\eta\down{1}{\ssty 2}}\up{-1}{\move{-1}\ssty (m)+}\move{-1}\bigl\}\move{2}\bigcup\\[1pt]
&\bigl\{
\ket{\phi\down{1}{\ssty 1}}\up{-1}{\move{-1}\ssty (\tilde{h})+},\ket{\psi\down{1}{\ssty 0}}\up{-1}{\move{-1}\ssty (\tilde{q})+},\ket{\psi\down{1}{\ssty 2}}\up{-1}{\move{-1}\ssty (\tilde{m})+}\bigl\}\down{-1}{\move{-1}\ssty x\neq\tilde{x}\in\{{0,1,2}\}},\\[5pt]
\mathcal{G}\up{1}{\ssty -}_{\ssty hqm}=
&\bigl\{
\ket{\psi\down{1}{\ssty 1}}\up{-1}{\move{-1}\ssty (h)-},\ket{\eta\down{1}{\ssty 0}}\up{-1}{\move{-1}\ssty (q)-},\ket{\eta\down{1}{\ssty 2}}\up{-1}{\move{-1}\ssty (m)-}\move{-1}\bigl\}\move{2}\bigcup\\[1pt]
&\bigl\{
\ket{\phi\down{1}{\ssty i}}\up{-1}{\move{-1}\ssty (g)-},\ket{\psi\down{1}{\ssty 0}}\up{-1}{\move{-1}\ssty (g)-},\ket{\psi\down{1}{\ssty 2}}\up{-1}{\move{-1}\ssty (g)-}\bigl\}\down{-1}{\move{-1}\ssty i,g\in\{0,1,2\}}
\end{array}
\end{equation*}
for each $h,q,m\in\{0,1,2\}$, where
\begin{equation}
\move{-3}
\begin{array}{ll}
%_{A\down{1.8}{\ssty k}A_{\overline{k+1}}A_{\overline{k+2}}}
\ket{\phi\down{1}{\ssty 0}}\up{-1}{\move{-1}\ssty (g)\pm} &\move{-5} =\ket{0}\ket{01\pm12},
\ket{\phi\down{0}{\ssty j}}\up{-1}{\move{-1}\ssty (g)\pm}= \ket{j}\ket{10\pm21},j=1,2,\\[1pt]
\ket{\psi\down{1}{\ssty i}}\up{-1}{\move{-1}\ssty (g)\pm}&\move{-5} = \ket{i}\ket{\overline{2i\move{-1}+\move{-1}2},i}+\omega_{\move{1}\ssty 2}^{\ssty\pm} \ket{\phi\down{1}{\ssty i}}\up{-1}{\move{-1}\ssty (g)+}\move{-5},\move{5}\overline{y} = y \move{-5}\mod 3, \\[1pt]
\ket{\eta\down{1}{\ssty i}}\up{-1}{\move{-1}\ssty (g)\pm} &\move{-5}= \ket{i}\ket{i,i}+\omega_{\move{1}\ssty 3}^{\ssty\pm}\ket{\psi\down{1}{\ssty i}}\up{-1}{\move{-1}\ssty (g)+}\move{-12}
\end{array}
\label{strong_nonlocal_basis}
\end{equation}

Here we commonly use the notation $\ket{\phi}\up{-1}{\move{-1}\ssty (g)}$ throughout this letter. The superscript ${}^{(g)}$ represents the cyclic rotation of the parties involved. Suppose $\ket{\phi}\up{-1}{\move{-1}\ssty (g)}=\ket{\alpha\beta\gamma}$ be a tripartite pure state with associated parties $A$,$B$ and $C$. Then $\ket{\phi}\up{-1}{\move{-1}\ssty (0)}=\ket{\alpha}_{\move{-1.5}\ssty A}\ket{\beta}_{\move{-1.5}\ssty B}\ket{\gamma}_{\move{-1.5}\ssty C}$, $\ket{\phi}\up{-1}{\move{-1}\ssty (1)}=\ket{\gamma}_{\move{-1.5}\ssty A}\ket{\alpha}_{\move{-1.5}\ssty B}\ket{\beta}_{\move{-1.5}\ssty C}$, and $\ket{\phi}\up{-1}{\move{-1}\ssty (2)}=\ket{\beta}_{\move{-1.5}\ssty A}\ket{\gamma}_{\move{-1.5}\ssty B}\ket{\alpha}_{\move{-1.5}\ssty C}$. 

The set $\mathcal{G}_{\ssty hqm}=\mathcal{G}\up{1}{\ssty +}_{\ssty hqm}\bigcup\mathcal{G}\up{1}{\ssty -}_{\ssty hqm}$ forms an orthogonal basis of biseparable states in $3\otimes3\otimes3$ Hilbert space. Note that  the state $\ket{\tau_{\ssty 3}^{\ssty 3}}$ is orthogonal to $\mathcal{G}\up{1}{\ssty -}_{\ssty hqm}$ but not to $\mathcal{G}\up{1}{\ssty +}_{\ssty hqm}$. So if we want to include the state $\ket{\tau_{\ssty 3}^{\ssty 3}}$ in $\mathcal{G}_{\ssty hqm}$, we need to disclude all the states of $\mathcal{G}\up{1}{\ssty +}_{\ssty hqm}$. In this way, we can define a new UBB $\mathcal{U}_{\ssty hqm}=\mathcal{G}\up{1}{\ssty -}_{\ssty hqm}\bigcup\ket{\tau_{\ssty 3}^{\ssty 3}}\}$, which is important in this work because it produce high dimensional genuinely entangled space ever exists. The proof of its unextendability is quite similar to the previous example. We only show that the $8$-dimensional subspace $\mathcal{GE}_{\ssty hqm}^{\ssty(8)}$ spanned by $\Omega_{\ssty 3}^{\ssty3}(\mathcal{G}\up{1}{\ssty +}_{\ssty hqm},\ket{\eta\down{1}{\ssty 0}}\up{-1}{\move{-1}\ssty (h)+})$ is genuinely entangled. For each bipartition, the dimension of the space spanned by the product-forming matrices and symmetrization matrices turns out to be $64$ which is exactly the square of the number of states in $\Omega_{\ssty 3}^{\ssty3}$. Therefore by Theorem \ref{sufficient_entangled},  $\mathcal{GE}_{\ssty hqm}^{\ssty (8)}$ is entangled in each bipartion and hence the set $\mathcal{U}_{\ssty hqm}$ follows Corollary \ref{sufficient_UBB}. 

$\mathcal{GE}_{\ssty hqm}^{\ssty (8)}$ holds particular significance as it represents the highest dimensional GES ever constructed from a UBB. A remarkable feature of this GES is its one-shot distillability across every bipartition, not merely for select states, but for every state within the subspace, regardless of purity. Any increase in the dimension of such a GES is therefore a nontrivial breakthrough in quantum information science. Notably, we have achieved an unprecedented expansion, increasing its dimension by three beyond the previously known maximal GES. The following lemma, introduced in \cite{Agrwal2019}, provides a direct proof of this one-shot distillability.

\begin{lemma}
Consider an $n$-dimensional subspace $S_{\ssty AB}$ of a bipartite Hilbert space $\mathbb{C}^{d_{\ssty  A}}\otimes\mathbb{C}^{d_{\ssty B}}.$ If the projector $\mathbb{P}_{\move{-1}\ssty AB}$ on $S_{\ssty  AB}$ satisfies the condition $\rank(\mathbb{P}_{\move{-1}\ssty AB})<\max\{\rank(\mathbb{P}_{\move{-1}\ssty A}),\rank(\mathbb{P}_{\move{-1}\ssty B})\}$, then all the rank-$n$ states supported on $\mathcal{S}_{\ssty AB}$ are $1$-distillable; where $\mathbb{P}_{\move{-1}\ssty A(B)}:=\Tr_{\ssty B(A)}(\mathbb{P}_{\move{-1}\ssty AB}).$
\end{lemma}

\begin{theorem} 
The subspace $\mathcal{GE}_{\ssty hqm}^{\ssty (8)}$ is $1$-distillable in every bipartition.
\end{theorem}

Proof: Consider an arbitrary projector $\mathbb{P}(n)$ of rank $n$ acting on
$\mathcal{GE}_{\ssty hqm}^{\ssty (8)}$,where $n\in[8]$. The following facts hold:
(i) Rank of bimarginal of the convex mixture of $n$ mutually orthogonal vectors cannot be less than that of the minimum number of states in $\Omega_{\ssty3}^{\ssty3}$ required to construct them.
(ii) To construct $n$ mutually orthogonal vectors in the subspace $\mathcal{GE}_{\ssty hqm}^{\ssty (8)}$ at least $n$ states from the set $\Omega_{\ssty 3}^{\ssty 3}$ are required.
(iii) Consider arbitrary $n$ number of states $\{\ket{\phi\down{1}{\ssty i}}\}_{\ssty i=0}^{\ssty n-1}$ from $\Omega_{\ssty 3}^{\ssty 3}$. For any such choices 
$\rank [\Tr_{\ssty\alpha}(\sum_{\ssty i=0}^{\ssty n-1}\ketbra{\phi\down{1}{\ssty i}}{\phi\down{1}{\ssty i}})]\geqslant n+1$, where $\alpha\in\{A,B,C\}$.

Therefore, bimarginals of an arbitrary projector $\mathbb{P}(n)$ have rank at least $n+1$ which assures distillability of the normalized projectors across every bipartition.\hfill$\ssty\blacksquare$\par

The distillability of the complementary space of $\mathcal{U}_{\ssty hqm}$ marks a significant achievement, something rarely realized in the constructions of UBBs. Every state in this complementary subspace can serve as a resource for quantum information tasks such as teleportation, entanglement swapping, and quantum key distribution. Distillability guarantees that pure entanglement can be extracted even from noisy states \cite{Bennett1996,Rains1999Rigorous,Rains1999Bound,Horodecki2001,Horodecki2009,Wang2016,Rozpedek2018,Regula2019,Chitambar2020,Biswas2023}, highlighting the practical utility of the subspace. In this sense, our construction not only advances the fundamental understanding of unextendible bases but also opens up operational advantages for real-world quantum technologies.

Building on these operational insights, we now shift our focus toward the theoretical foundations of nonlocality arising from local state discrimination \cite{NonlocalityIndistinguishability}. While the unextendibility of $\mathcal{U}_{\ssty hqm}$ inherently guarantees its nonlocal nature, we take this a step further by showing that it exhibits strong nonlocality. To this end, we introduce the following theorem, which provides a sufficient condition for a set to be locally irreducible that directly leads to the condition of strong nonlocality.

The application of this theorem requires a careful matrix analysis. We begin by considering the bases of $  A$ and $ B$ as described in Eq. (\ref{general_form}) are orthonormal. For each $i,j(\neq i)\in[n]$, we define $d_{\ssty A}\cross d_{\ssty A}$ matrices $\Pi_{\ssty ij}^{\ssty A}$ as: 
\begin{equation}
(\Pi_{\ssty ij}^{\ssty    A})\down{1}{\ssty kp}=\sum_{\ssty l=0}^{\ssty d_{ B}-1} \overline{a_{\ssty kl}^{\ssty(i)}}a_{\ssty pl}^{\ssty(j)}
\end{equation}
where $ k,p=[d_{\ssty A}]$. These are total $d_{\ssty A}(d_{\ssty A}-1)$ matrices, said to be reduced feature matrices over $A$. By analyzing these matrices, we establish a connection to the condition of local irreducibility, which is intrinsically linked to the concept of strong nonlocality. 

\begin{theorem}
In a discrimination task of orthogonal pure states in $\mathcal{S}$, Alice goes first with nontrivial OPLM, if and only if all the reduced feature matrices over $A$ span the subspace of $M(d_{\ssty A},\mathbb{C})$ with dimension less than $(d_{\ssty A}-1)$.
\label{nontrivial_condition}
\end{theorem}

$\vspace{-10pt}$

Proof: Let us consider the measurement operator chosen by Alice can be expressed as  $E=\sum_{\ssty kp} e_{\ssty kp}\ketbra{k}{p}, k,p\in[d_{\ssty  A}]$. Therefore, 
$\vspace{-10pt}$
\begin{equation}
\move{-3}
\begin{array}{rl}
\bra{\psi\down{1}{\ssty i}}E_{\ssty  A}\otimes\mathbb{I}_{\ssty B}\ket{\psi\down{0}{\ssty j}}\move{-4}
&=\underset{\ssty kl}{\sum} \overline{a_{\ssty kl}^{\ssty (i)}}\bra{kl}E_{\ssty  A}\otimes\mathbb{I}_{\ssty B}\ket{\psi\down{0}{\ssty j}}\\[2pt]
&=\underset{\ssty kl}{\sum} \overline{a_{\ssty kl}^{\ssty (i)}}(\sum_{\ssty p} e_{\ssty kp}\braket{ pl}{\psi\down{0}{\ssty j}})
=\underset{\ssty kpl}{\sum} \overline{a_{\ssty kl}^{\ssty (i)}}a_{\ssty pl}^{\ssty (j)} e_{\ssty kp}
\end{array}
\label{postmeasurement_form}
\end{equation}

Since the measurement chosen by Alice is OPLM, then for each $i,j(\neq i)\in[n]$,
\begin{equation}
\begin{array}{lll}
&\bra{\psi\down{1}{\ssty i}}E_{\ssty  A}\otimes\mathbb{I}_{\ssty B}\ket{\psi\down{0}{\ssty j}}=0\\[2pt]
\implies&\underset{\ssty kp}{\sum} (\underset{\ssty l}{\sum}\overline{a_{\ssty kl}^{\ssty (i)}}a_{\ssty pl}^{\ssty (j)}) e_{\ssty kp}=0,
\text{ i.e. },\underset{\ssty kp}{\sum} e_{\ssty kp}(\Pi_{\ssty ij}^{\ssty    A})\down{1}{\ssty kp}=0
\end{array}
\label{orthogonal_postmeasurement}
\end{equation}

$\mathbb{I}_{\ssty A}$ is solution of the Eqs. (\ref{orthogonal_postmeasurement}) since $\braket{\psi\down{1}{\ssty i}}{\psi\down{0}{\ssty j}}=0$ for $i\neq j$. Alice can't go first with nontrivial OPLM if $\mathbb{I}$ is the only solution of the Eqs. (\ref{orthogonal_postmeasurement}). In otherwords the matrices $\{\Pi_{\ssty ij}^{\ssty    A}\}\down{1}{\ssty i,j}$ spans the complement of $ \mathbb{I}_{\ssty A}$.

Conversely, if possible let there exist $\tilde{E}\not\propto\mathbb{I}$ which satisfies the Eqs. (\ref{orthogonal_postmeasurement}). Suppose $\tilde{E}=[\tilde{e}_{\ssty kp}]\down{1}{\ssty k,p}$ is not a Hermitian matrix, then 
\begin{equation}
\begin{array}{ll}
&\bra{\psi\down{0}{\ssty j}}\tilde{E}^{\dagger}_{\ssty A}\otimes\mathbb{I}_{\ssty B}\ket{\psi\down{1}{\ssty i}}\\[2pt]
=&\underset{\ssty pkl}{\sum} \overline{a_{\ssty pl}^{\ssty (j)}}a_{\ssty kl}^{\ssty (i)} \overline{\tilde{e}_{\ssty kp}}=\overline{\underset{\ssty kpl}{\sum} \overline{a_{\ssty kl}^{\ssty (i)}}a_{\ssty pl}^{\ssty  (j)}\tilde{e}_{\ssty kp}}=0\quad\text{[by (\ref{orthogonal_postmeasurement})]}
\end{array}
\label{bipartite_form}
\end{equation}

That means $\tilde{E}^{\dagger}$ is a solution of the Eqs. (\ref{orthogonal_postmeasurement}). Therefore $\tilde{E}'=(\tilde{E}+\tilde{E}^{\dagger})$ and $\tilde{E}''=\mathrm{i}(\tilde{E}-\tilde{E}^{\dagger})$, $\mathrm{i}=\sqrt{-1}$ are Hermitian solution of the Eqs. (\ref{orthogonal_postmeasurement}), not both proportional to $\mathbb{I}$.
Therefore there always exist a Hermitian matrix $E^{\ast}\not\propto\mathbb{I}$ as a solution of the Eqs. (\ref{orthogonal_postmeasurement}). Consequently $\{(\mathbb{I}+E^{\ast}/\lambda^{\ast})/2,(\mathbb{I}-E^{\ast}/\lambda^{\ast})/2\}$  is the nontrivial OPLM can be chosen by Alice, where $\lambda^{\ast}$ is the dominant eigenvalue of $E^{\ast}$. This completes the theorem.\hfill$\ssty\blacksquare$\par

The contrapositive of this theorem naturally establishes a no-go condition within our framework: if the stated requirement fails for a given set of bipartite states, irrespective of party choices, then no party can perform a nontrivial measurement. As a result, no state can be conclusively eliminated, making the set locally irreducible. Extending this principle to multipartite systems, a set of orthogonal states exhibits strong nonlocality if it satisfies this no-go condition across every bipartition. This insight sets the stage for proving the next theorem.

\begin{theorem}
The set $\mathcal{U}_{\ssty hqm}$ is strongly nonlocal for each $h,q,m=0,1,2$.
\end{theorem}
Consider the bipartition $AB|C$. We find that the reduced feature matrices over the subsystem $AB$ span a subspace of dimension 80, which is precisely one less than the full dimension of $ M(9,\mathbb{C})$. As a result, neither Alice nor Bob can perform any nontrivial measurements locally as well as jointly. Consequently, the set remains irreducible both locally and across the $AB|C$ bipartition. A similar situation arises in the other two bipartitions as well. Hence, it becomes impossible for any party or group of parties to eliminate even a single state from the set using orthogonality-preserving measurements. This complete irreducibility across all partitions is a manifestation of strong nonlocality.

Our construction of strongly nonlocal UBBs resolves an open problem posed by Agrawal \textit{et al.} \cite{Agrwal2019}. The UBBs produce genuinely entangled subspaces, the states of which are distillable across every bipartition. This result pushes the known boundaries of distillable GES, offering new insights into the interplay between quantum nonlocality, entanglement structure, and resource theory. A related investigation was carried out in our previous work \cite{Atanu2024}, but here, we achieve a notable breakthrough by increasing the dimension of the GES by $3$.  This advancement marks a pivotal step in quantum information science, emphasizing the intricate interplay between unextendibility and strong nonlocality. A key challenge lies in proving both the strong nonlocality of the UBB and the genuineness of its complementary space. We introduce sufficient conditions for both, which not only validate our construction but also provide a framework for analyzing existing UBBs. Notably, our genuineness condition serves as a criterion for proving the unextendibility of any preexisting UBB, while our strong nonlocality condition generalizes to capture strong nonlocality in any known strongly nonlocal set. Additionally, we establish that every UBB is locally indistinguishable, though not necessarily strongly nonlocal, challenging previous misconceptions and refining our understanding of multipartite quantum systems.

Our constructional symmetry allows for a natural extension of this idea to tripartite and higher dimensional multipartite systems. However, our primary motivation for expanding the dimension of a GES with distillable entanglement has already been achieved in the presented example. While the explicit construction of such UBBs follows a similar approach, their underlying physical properties can vary significantly. It will be interesting to distinguish such UBBs, relying on the information contained within them. Several fundamental problems remain open. The minimum cardinality of a UBB and its optimal construction are yet to be determined. It is known that the maximum dimension of a GES is $d^3-d^2-d+1$ in the $d\otimes d\otimes d$ system. Therefore, the minimum possible cardinality of a UBB is $d^2+d-1$. Additionally, a long-standing open question remains unresolved: Does there exist a UPB whose complementary space consists entirely of genuinely entangled states? In our terms, the existence of a UPB which is also a UBB remains unknown. Moreover, in \cite{Watrous2005, Jiang2009}, the authors introduced the notion of an indistinguishable subspace, defined as a subspace where every orthonormal basis is locally indistinguishable. In \cite{Runyo2010}, Duan \textit{et al.} proved that the subspace spanned by the three-qubit UPB is indistinguishable under LOCC. Motivated by these results, it is natural to expect that every UBB constructed in this work forms an indistinguishable subspace, and that their complementary subspaces also exhibit this property. However, a formal proof of this remains a challenging open problem. In this context, the problem of finding a strongly nonlocal subspace or at least a genuinely nonlocal subspace also remains open. It is known that any $2\otimes n$ subspace admits a basis that is locally distinguishable \cite{Yu2011}. Consequently, no genuinely nonlocal subspace can exist in any multipartite system that includes a local dimension of $2$. Therefore, an intriguing direction is to investigate whether a UBB or its complementary subspace can span a strongly or genuinely nonlocal subspace in higher-dimensional systems. Addressing these challenges will advance our understanding of quantum nonlocality, unextendibility, and entanglement structure, with profound implications for quantum cryptography and quantum communication.

\textit{\textbf{Acknowledgement}}---The author S. Bera acknowledges the support from CSIR, India. The author I. Biswas acknowledges the support from UGC, India. The authors I. Chattopadhyay and D. Sarkar acknowledge the DST-FIST India.

\begin{center}	
\bf{Appendix}
\end{center}
%\onecolumn
\section{Preliminaries}
\begin{definition}
In mathematics, especially in linear algebra and matrix theory, the vectorization of a matrix is a linear transformation that converts the matrix into a vector. Specifically, the vectorization of a $m \times n$ matrix $A$, denoted $\operatorname{vec}(A)$, is the $m n \times 1$ column vector obtained by stacking the columns of the matrix $A$ on top of one another:

$$
\operatorname{vec}(A)=\left[a_{\ssty11}, \ldots, a_{m\ssty1}, a_{\ssty12}, \ldots, a_{m\ssty2}, \ldots, a_{{\ssty1}n}, \ldots, a_{mn}\right]^{\mathrm{T}}
$$

Here, $a_{\ssty ij}$ represents the element in the $i$-th row and $j$-th column of $A$, and the superscript ${ }^{\mathrm{T}}$ denotes the transpose. 
\end{definition}

Let $A_{1}, A_2,\ldots,\text{and }A_k$ be $k$ matrices of order $m \times n$. Let a matrix M of order $mn \times k$ whose $i$-th column is the vectorization of $A_i,i=1,2,\ldots,k$. Vectorization expresses, through coordinates, the isomorphism $\mathbf{C}^{m \times n}:=\mathbf{C}^m \otimes \mathbf{C}^n \cong \mathbf{C}^{m n}$ between the vector space $V$ spanned by the matrices $\{A_i\}_{i}$ and the columnspace of $M$. Therefore, the dimension of $V$ is equal to the rank of $M$. The matrix $M$ is defined to be the column-wise vectorization map of the matrices $\{A_i\}_{i}$.
\pagebreak

\section{Unextendibility of $\mathcal{U}$}\label{Algorithm genuinity}

In this section, we compute each of the product-forming and symmetrization matrices of $\Omega_{\ssty 2}^{\ssty 3}(\mathcal{G}\up{1}{\ssty +},\ket{\eta}^{\move{-1}\ssty+})$ corresponding to different bipartitions. We then examine their structure to demonstrate that the space they span contains only genuinely entangled states.

The explicit form of the states of $\mathcal{G}\up{1}{\ssty +}$ and $\Omega_{\ssty 2}^{\ssty 3}(\mathcal{G}\up{1}{\ssty +},\ket{\eta}^{\move{-1}\ssty+})$ are written as
\begin{equation}
\ket{\eta}^{\move{-1}\ssty+} = \ket{01+10}\ket{1},
\ket{\psi\down{1}{ 0}}^{\move{-1}\ssty+} = \ket{0}\ket{00+10+01},\ket{\psi\down{1}{ 1}}^{\move{-1}\ssty+} = \ket{1}\ket{00+10+11},\text{ and}
\label{genuinely_local_basis}
\end{equation}
\begin{equation}
\ket{\eta\down{1}{ 0}}= 2\ket{0}\ket{00+10+01}-3\ket{01+10}\ket{1},
\ket{\eta\down{1}{ 1}}= 2\ket{1}\ket{00+10+11}-3\ket{01+10}\ket{1}
\end{equation}
respectively.

Consider the $A|BC$ partition. Then the states of $\Omega_{\ssty 2}^{\ssty 3}(\mathcal{G}\up{1}{\ssty +},\ket{\eta}^{\move{-1}\ssty+})$ can be expressed as 
\begin{equation}
\ket{\psi\down{1}{ 0}}= 2\ket{0}\ket{0+1+2}-3\ket{03+11},
\ket{\psi\down{1}{ 1}}= 2\ket{1}\ket{0+2+3}-3\ket{03+11}
\end{equation}
in the basis $\ket{00}_{\move{-1.5}\ssty BC}\equiv\ket{0}_{\move{-1.5}\ssty BC}$,$\ket{01}_{\move{-1.5}\ssty BC}\equiv\ket{1}_{\move{-1.5}\ssty BC}$,$\ket{10}_{\move{-1.5}\ssty BC}\equiv\ket{2}_{\move{-1.5}\ssty BC}$, and $\ket{11}_{\move{-1.5}\ssty BC}\equiv\ket{3}_{\move{-1.5}\ssty BC}$

%\begin{table}[h]
%\caption{\label{genuinely_local_basis_table}$\ket{\psi\down{1}{i}}_{\move{-1.5}\ssty AB}=\sum_{\ssty kl} a_{\ssty kl}^{\ssty (i)}\ket{kl}_{\move{-1.5}\ssty AB}$}
%\begin{ruledtabular}
%\begin{tabular}{|c|c|cccc|}
%\textbf{states($\ket{\psi\down{1}{ i}}$)} 		& $a_{kl}^{ (i)}$ 	& l=0 	& l=1 	& l=2		& l=3 \\
%\hline
%i=0 								& k=0			&2		& 2 		&2		&-3\\
% 									& k=1			&0		& -3 		&0		&0\\ 
%\hline
% 					
%i=1 								& k=0			&0		&  0		&0		&-3\\
% 									& k=1			&2		& -3 		&2		&2
%\end{tabular}
%\end{ruledtabular}
%\end{table}

\begin{table}[h]
    \centering
    \caption{\label{genuinely_local_basis_table_I}$a_{kl}^{(i)}\left(\ket{\psi\down{1}{i}}_{\move{-1.5}\ssty AB}=\sum_{kl} a_{kl}^{(i)}\ket{kl}_{AB}\right)$}
    \begin{tabular}{c|c|cccc}
        \toprule
        $i$ & $k \setminus l$ & 0 & 1 & 2 & 3 \\
        \midrule
        0& 0 & 2 & 2 & 2 & -3 \\
          & 1 & 0 & -3 & 0 & 0 \\
        \midrule
        1& 0 & 0 & 0 & 0 & -3 \\
          & 1 & 2 & -3 & 2 & 2 \\
        \bottomrule
    \end{tabular}
\end{table}

The related product forming matrices of $\Omega_{\ssty 2}^{\ssty 3}(\mathcal{G}\up{1}{\ssty +},\ket{\eta}^{\move{-1}\ssty+})$ are

\begin{equation}
\begin{array}{l}
\Lambda_{\ssty 01|01}^{\ssty    A|BC}=
\begin{bmatrix}
-6& -10\\
0 & 0
\end{bmatrix},
\Lambda_{\ssty 01|02}^{\ssty    A|BC}=
\begin{bmatrix}
0& 0\\
0 & 0
\end{bmatrix},
\Lambda_{\ssty 01|03}^{\ssty    A|BC} =
\begin{bmatrix}
0& 10\\
0 & 6
\end{bmatrix},\\[15pt]
\Lambda_{\ssty 01|12}^{\ssty    A|BC}=
\begin{bmatrix}
6& 10\\
0 & 0
\end{bmatrix},
\Lambda_{\ssty 01|13}^{\ssty    A|BC}=
\begin{bmatrix}
-9& -5\\
-9 & -9
\end{bmatrix},
\Lambda_{\ssty 01|23}^{\ssty    A|BC}=
\begin{bmatrix}
0& 10\\
0 & 6
\end{bmatrix}
\end{array}
\label{product_forming_matrices}
\end{equation}

The related symmetrization matrix of $\Omega_{\ssty 2}^{\ssty 3}(\mathcal{G}\up{1}{\ssty +},\ket{\eta}^{\move{-1}\ssty+})$ is 

\begin{equation}
\Gamma_{\move{-2}\ssty 01}=
\begin{bmatrix}
0 & 1\\
-1& 0
\end{bmatrix}
\label{symmetrization_matrices}
\end{equation}

The expression for the symmetrization matrices for an arbitrary set of the same number of states is unique, considering any bipartition. But, it is not the case for the product forming matrices because it can be changed depending on the basis chosen.

The columnwise vectorization map of the matrices in (\ref{product_forming_matrices},\ref{symmetrization_matrices}) is 
\begin{equation}
M^{\ssty    A|BC}=
\begin{bmatrix}
6 	& 0 	& 0 	& 6 	& -9 	& 0 	& 0\\
0 	& 0 	& 0 	& 0 	& -9 	& 0 	& -1\\
-10& 0 	& 10 	& 10 	& -5 	& 10 	& 1\\
0 	& 0 	& 6 	& 0 	& 9 	& 6 	& 0\\

\end{bmatrix}
\label{vectorization_map_A_BC}
\end{equation}

The rank of $M^{\ssty    A|BC}$ is $4$. Then the dimension of the above matrices is 4.

Similarly, we can evaluate the column-wise vectorization map for the rest of two bipartitions, which are as follows:

\begin{equation}
\move{26}M^{\ssty    B|CA}=
\begin{bmatrix}
0 	& -10 & 6 	& 0 	& 0	& -9 	& 0\\
-4 	& 0 	& 6 	& -6 	& 0 	& -9 	& -1\\
4 	& -6 	& 4 	& -4 	& 6 	& -5 	& 1\\
0 	& 0 	& 0 	& -6 	& 10 	& -9 	& 0\\

\end{bmatrix}, \text{ and}
\label{vectorization_map_B_CA}
\end{equation}

\begin{equation}
M^{\ssty    C|AB}=
\begin{bmatrix}
-10 	& -6	& 0 	& -6 	& 0	& 0 	& 0\\
0	& -4 	& -4	& 6 	& 6 	& 6 	& -1\\
-6 	& -6 	& 4 	& -6 	& 4 	& 0 	& 1\\
0 	& 0 	& 0 	& 6 	& 6 	& 10 	& 0\\

\end{bmatrix}
\label{vectorization_map_B_CA}
\end{equation}
Both the matrices have rank $4$.

\section{Proof of Theorem 3}
The explicit form of the states of $\mathcal{U}$ is written as
\begin{equation}
\begin{array}{c}
\ket{\phi\down{1}{ 0}}^{\move{-1}\ssty-} \move{-3}= \ket{0}\ket{00-10},
\ket{\phi\down{1}{ 1}}^{\move{-1}\ssty-} \move{-3}= \ket{1}\ket{00-10},
\ket{\eta}^{\move{-1}\ssty-} \move{-3}= \ket{01-10}\ket{1},\\
\ket{\psi\down{1}{ 0}}^{\move{-1}\ssty-} = \ket{0}(\ket{00}+\ket{10}-2\ket{01}),\ket{\psi\down{1}{ 1}}^{\move{-1}\ssty-} = \ket{1}(\ket{00}+\ket{10}-2\ket{11}),\\
\ket{\tau_{\ssty 2}^{\ssty 3}}=\ket{0+1}\ket{0+1}\ket{0+1}
\end{array}
\label{genuinely_local_basis}
\end{equation}

Consider the $A|BC$ partition. Then the states of $\mathcal{U}$ can be expressed as 
\begin{equation}
\begin{array}{c}
\ket{\phi\down{1}{ 0}}^{\move{-1}\ssty-} \move{-3}= \ket{0}\ket{0-2},
\ket{\phi\down{1}{ 1}}^{\move{-1}\ssty-} \move{-3}= \ket{1}\ket{0-2},
\ket{\eta}^{\move{-1}\ssty-} \move{-3}= \ket{03-11},\\
\ket{\psi\down{1}{ 0}}^{\move{-1}\ssty-} = \ket{0}(\ket{0}+\ket{2}-2\ket{1}),\ket{\psi\down{1}{ 1}}^{\move{-1}\ssty-} = \ket{1}(\ket{0}+\ket{2}-2\ket{3}),\\
\ket{\tau_{\ssty 2}^{\ssty 3}}=\ket{0+1}\ket{0+1+2+3}
\end{array}
\label{genuinely_local_basis_A_BC}
\end{equation}

Using the following protocol, it can be shown that the states in (\ref{genuinely_local_basis_A_BC}) are locally distinguishable.

\begin{forest}
  for tree={
    edge={draw, ->}, % Draw edges with arrows
    l sep=30pt       % Adjust the spacing between levels
  }
  [$\mathcal{U}$
  [Bob and Charlie
    [
    $\begin{aligned}
    &\ket{0}\ket{0-2}\text{,}\ket{1}\ket{0-2} \\
    &\ket{0}(\ket{0}+\ket{2}-2\ket{1}) \text{,}\ket{1}(\ket{0}+\ket{2}-2\ket{3}) \\
    &\ket{0}(\ket{0+1+2}-2\ket{3})-\ket{1}(\ket{0+2+3}-2\ket{1})
    \end{aligned}$,
     edge label={node[midway, sloped, below]{$\mathbb{I}-\ketbra{\tau_{\ssty 2}^{\ssty 2}}{\tau_{\ssty 2}^{\ssty 2}}$}}
    [Alice
      [$\begin{aligned}
    &\ket{0}\ket{0-2}\\
    &\ket{0}(\ket{0}+\ket{2}-2\ket{1})\\
    &\ket{0}(\ket{0+1+2}-2\ket{3})
    \end{aligned}$,
      edge label={node[midway, sloped, above]{$\ketbra{0}{0}$}}]
      [$\begin{aligned}
    &\ket{1}\ket{0-2}\\
    &\ket{1}(\ket{0}+\ket{2}-2\ket{3})\\
    &\ket{1}(\ket{0+2+3}-2\ket{1})
    \end{aligned}$,
      edge label={node[midway, sloped, above]{$\ketbra{1}{1}$}}]
    ]]
    [$\ket{0\pm1}\ket{0+1+2+3} $, 
      edge label={node[midway, sloped, below]{$\ketbra{\tau_{\ssty 2}^{\hspace{1pt}\ssty 2}}{\tau_{\ssty 2}^{\ssty 2}}$}}
    ]
    ]
  ]
\end{forest}

The states at each end of the tree are locally distinguishable, hence the states in (\ref{genuinely_local_basis_A_BC}).

\section{Unextendibility of $\mathcal{U}_{\ssty hqm}$}

At the very first, we will discuss the construction of these sets using the position index of matrices. Consider a square matrix of order $3$ and think of the position of the matrix element as mutually orthogonal bipartite product states. Our construction produces different UBBs associated with any off-diagonal position. Consider such a position $(i,j)$ and eliminate the corresponding row and column. Using the column number $j$ and the diagonal position of the remaining matrix, here denoted as $M_{\ssty ij}$, we define the following states, orthogonal to the stopper state $\ket{\tau_{\ssty 3}^{\ssty 3}}$ of the associated system:
\begin{equation}
\ket{\phi\down{2}{\ssty ij}}\up{-1}{\move{-1}\ssty (g)-}=\ket{j}\ket{\alpha\down{1}{\ssty ij}^{\ssty -}},
\ket{\psi\down{2}{\ssty ij}}\up{-1}{\move{-1}\ssty (g)-}=\ket{j}\ket{\beta\down{1}{\ssty ij}^{\ssty -}}
\end{equation}
where $\ket{\alpha\down{1}{\ssty ij}^{\ssty -}}$,$\ket{\beta\down{1}{\ssty ij}^{\ssty -}}$ are the bipartite states of two qubit system. These states, along with $\ket{\alpha\down{1}{\ssty ij}^{\ssty +}}$ , form a mutually orthogonal set, where each state is a superposition of $\ket{ij}$ and the diagonal states of $M_{\ssty ij}$. The superscript $^{\ssty +}$  indicates an equal superposition. Here $g =0,1,2$ is associated with the cyclic rotations of parties $ABC$, $BCA$, and $CAB$ respectively.

Similarly, the UBB includes the states 
\begin{equation}
\ket{\phi\down{2}{\ssty ji}}\up{-1}{\move{-1}\ssty (g)-}=\ket{i}\ket{\alpha\down{1}{\ssty ji}^{\ssty -}},
\ket{\psi\down{2}{\ssty ji}}\up{-1}{\move{-1}\ssty (g)-}=\ket{i}\ket{\beta\down{1}{\ssty ji}^{\ssty -}}
\end{equation}

Now for $k\neq i,j$, the UBB includes the following states: 
\begin{equation}
\ket{\eta\down{2}{\ssty ij}}\up{-1}{\move{-1}\ssty (g)-}=\ket{k}\ket{\gamma\down{1}{\ssty ij}^{\ssty -}}
\end{equation}
 where $\ket{\gamma\down{1}{\ssty ij}^{\ssty -}}$ is a two qubit pure state. The state, along with $\ket{\gamma\down{1}{\ssty ij}^{\ssty +}}$, form a mutually orthogonal set, where each state is a superposition of the diagonal states of $M_{\ssty ij}$.

%Now we are going to add $3$ asymmetric states along with three-qutrit stopper $\ket{\tau_{\ssty 3}^{\ssty 3}}$ to complete the UBB.

So far, we have defined 15 mutually orthogonal pure states, including all possible cyclic rotations of the parties. We can see that they included all the positional states of the actual matrix except the diagonal ones. Then for the rotations $h,q,m$, let us define

\begin{equation}
\ket{\tilde{\phi}\down{2}{\ssty ij}}\up{-1}{\move{-1}\ssty (h)}=\ket{j}\ket{\tilde{\alpha}\down{1}{\ssty ij}^{\ssty -}},
\ket{\tilde{\phi}\down{2}{\ssty ji}}\up{-1}{\move{-1}\ssty (q)-}=\ket{i}\ket{\tilde{\alpha}\down{1}{\ssty ji}^{\ssty -}},
\ket{\tilde{\eta}\down{2}{\ssty ij}}\up{-1}{\move{-1}\ssty (m)-}=\ket{k}\ket{\tilde{\gamma}\down{1}{\ssty ij}^{\ssty -}}
\end{equation}
 where $\ket{\tilde{\alpha}\down{1}{\ssty ij}^{\ssty -}}$, $\ket{\tilde{\alpha}\down{1}{\ssty ji}^{\ssty -}}$, and $\ket{\tilde{\gamma}\down{1}{\ssty ij}^{\ssty -}}$ are sperposition of $\ket{jj}$ and $\ket{\alpha\down{1}{\ssty ij}^{\ssty +}}$; $\ket{ii}$ and $\ket{\alpha\down{1}{\ssty ji}^{\ssty +}}$; $\ket{kk}$ and $\ket{\gamma\down{1}{\ssty ij}^{\ssty +}}$ respectively, such that they became orthogonal to $\ket{\tau_{\ssty 3}^{\ssty 3}}$. We claim the above define $18$ states along with the stopper $\ket{\tau_{\ssty 3}^{\ssty 3}}$ forms an UBB. The set $\mathcal{U}_{\ssty hqm}$ can be similarly constructed if we choose the position $(2,0)$.

As we can see, the explicit structures of each $\mathcal{U}_{\ssty hqm}$ are similar. So it is enough to prove their unextendibility for $h=0$,  $q=0$, and $m=0$. The states of the set $\mathcal{U}_{\ssty 000}$ can be expressed as: 

\begin{equation}
\begin{array}{l}
\ket{\phi\down{1}{\ssty 0}}\up{-1}{\move{-1}\ssty (g)-} =\ket{0}\ket{01-12},
\ket{\phi\down{0}{\ssty 1}}\up{-1}{\move{-1}\ssty (g)-}= \ket{1}\ket{10-21},
\ket{\phi\down{0}{\ssty 2}}\up{-1}{\move{-1}\ssty (g)-}= \ket{2}\ket{10-21}\\[2pt]

\ket{\psi\down{1}{\ssty 0}}\up{-1}{\move{-1}\ssty (g)-}= \ket{0}(\ket{01+12}-2\ket{20}),
 \ket{\psi\down{1}{\ssty 2}}\up{-1}{\move{-1}\ssty (g)-}= \ket{2}(\ket{10+21}-2\ket{02}),\\[2pt]
 
\ket{\eta\down{1}{\ssty 0}}\up{-1}{\move{-1}\ssty (0)-} = \ket{0}_{\move{-1.5}\ssty A}(\ket{01+12+20}-3\ket{00})_{\move{-1.5}\ssty BC},
\ket{\eta\down{1}{\ssty 2}}\up{-1}{\move{-1}\ssty (0)-} = \ket{2}_{\move{-1.5}\ssty A}(\ket{10+21+02}-3\ket{22})_{\move{-1.5}\ssty BC}\\[2pt]

 \ket{\psi\down{1}{\ssty 1}}\up{-1}{\move{-1}\ssty (0)-}= \ket{1}_{\move{-1.5}\ssty A}(\ket{10+21}-2\ket{11})_{\move{-1.5}\ssty BC},\ket{\tau_{\ssty 3}^{\ssty 3}}=\ket{0+1+2}_{\move{-1.5}\ssty A}\ket{0+1+2}_{\move{-1.5}\ssty B}\ket{0+1+2}_{\move{-1.5}\ssty C}
\end{array}
\label{strong_nonlocal_basis}
\end{equation}

\begin{figure}[htp]
    \centering
    \includegraphics[scale=.5]{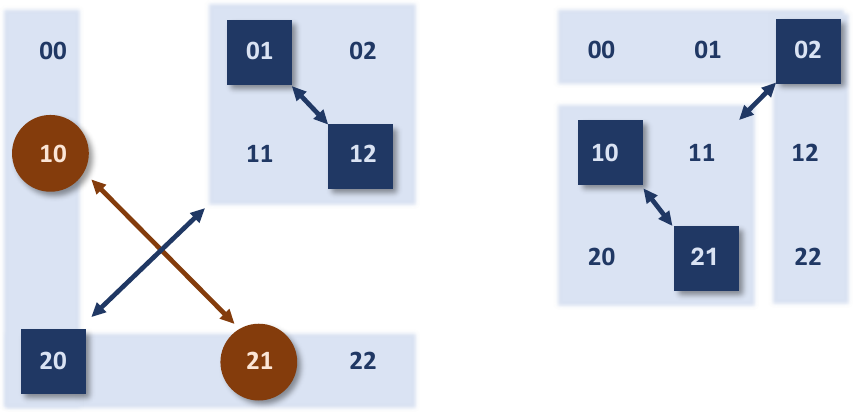}
     \caption{Pictorial representation of UBB, $\mathcal{U}_{\ssty 000}$ produced when considering the (0,2 ) position of a matrix of order $3$ and the rotation $h=0$, $q=0$, and $m=0$. }
    \label{fig:UBB}
\end{figure}

Therefore $\mathcal{U}_{\ssty 000}$ consists of total 19 states. The corresponding complementary space is the span of $\Omega_{\ssty 3}^{\ssty3}(\mathcal{G}\up{1}{\ssty +}_{\ssty 000},\ket{\eta\down{1}{\ssty 0}}\up{-1}{\move{-1}\ssty (0)+})$, consisting the following pure states:

\begin{equation}
\begin{array}{l}
 \ket{\psi\down{1}{\ssty 0}}= 4\ket{1}\ket{10+11+21}-3\ket{0}\ket{00+01+12+20},\\[2pt]
 \ket{\psi\down{1}{\ssty 1}}= \ket{2}\ket{02+10+21+22}-\ket{0}\ket{00+01+12+20},\\[2pt]
  \ket{\psi\down{1}{\ssty 2}}= \ket{0}(4\ket{02}-3\ket{00+01+12+20})+4\ket{100+201},\\[2pt]
   \ket{\psi\down{1}{\ssty 3}}= \ket{0}(4\ket{10}-3\ket{00+01+12+20})+4\ket{120+200},\\[2pt]
    \ket{\psi\down{1}{\ssty 4}}= \ket{0}(2\ket{11}-\ket{00+01+12+20})+2\ket{112},\\[2pt]
     \ket{\psi\down{1}{\ssty 5}}= 2\ket{101+211}-\ket{0}\ket{00+01+12+20},\\[2pt]
 \ket{\psi\down{1}{\ssty 6}}= \ket{0}(4\ket{21}-3\ket{00+01+12+20})+4\ket{122+220},\\[2pt]
  \ket{\psi\down{1}{\ssty 7}}= \ket{0}(4\ket{22}-3\ket{00+01+12+20})+4\ket{102+212},\\[2pt]
\end{array}
%\label{strong_nonlocal_basis}
\end{equation}

Consider the $A|BC$ partition. Then the states of $\Omega_{\ssty 3}^{\ssty3}(\mathcal{G}\up{1}{\ssty +}_{\ssty 000},\ket{\eta\down{1}{\ssty 0}}\up{-1}{\move{-1}\ssty (0)+})$ can be expressed as 
\begin{equation}
\begin{array}{l}
 \ket{\psi\down{1}{\ssty 0}}= 4\ket{1}\ket{3+4+7}-3\ket{0}\ket{0+1+5+6},\\[2pt]
 
 \ket{\psi\down{1}{\ssty 1}}= \ket{2}\ket{2+3+7+8}-\ket{0}\ket{0+1+5+6},\\[2pt]
  \ket{\psi\down{1}{\ssty 2}}= \ket{0}(4\ket{2}-3\ket{0+1+5+6})+4\ket{10+21},\\[2pt]
   \ket{\psi\down{1}{\ssty 3}}= \ket{0}(4\ket{3}-3\ket{0+1+5+6})+4\ket{16+20},\\[2pt]
    \ket{\psi\down{1}{\ssty 4}}= \ket{0}(2\ket{4}-\ket{0+1+5+6})+2\ket{15},\\[2pt]
     \ket{\psi\down{1}{\ssty 5}}= 2\ket{11+24}-\ket{0}\ket{0+1+5+6},\\[2pt]
 \ket{\psi\down{1}{\ssty 6}}= \ket{0}(4\ket{7}-3\ket{0+1+5+6})+4\ket{18+26},\\[2pt]
  \ket{\psi\down{1}{\ssty 7}}= \ket{0}(4\ket{8}-3\ket{0+1+5+6})+4\ket{12+25},\\[2pt]
\end{array}
%\label{strong_nonlocal_basis}
\end{equation}
in the basis $\ket{00}_{\move{-1.5}\ssty BC}\equiv\ket{0}_{\move{-1.5}\ssty BC}$,$\ket{01}_{\move{-1.5}\ssty BC}\equiv\ket{1}_{\move{-1.5}\ssty BC}$,$\ket{02}_{\move{-1.5}\ssty BC}\equiv\ket{2}_{\move{-1.5}\ssty BC}$, $\ket{10}_{\move{-1.5}\ssty BC}\equiv\ket{3}_{\move{-1.5}\ssty BC}$, $\ket{11}_{\move{-1.5}\ssty BC}\equiv\ket{4}_{\move{-1.5}\ssty BC}$, $\ket{12}_{\move{-1.5}\ssty BC}\equiv\ket{5}_{\move{-1.5}\ssty BC}$, $\ket{20}_{\move{-1.5}\ssty BC}\equiv\ket{6}_{\move{-1.5}\ssty BC}$, $\ket{21}_{\move{-1.5}\ssty BC}\equiv\ket{7}_{\move{-1.5}\ssty BC}$, and $\ket{22}_{\move{-1.5}\ssty BC}\equiv\ket{8}_{\move{-1.5}\ssty BC}$.

\begin{table}[h]
    \caption{\label{genuinely_local_basis_table_II}$a_{kl}^{(i)}\left(\ket{\psi\down{1}{i}}_{\move{-1.5}\ssty AB}=\sum_{kl} a_{kl}^{(i)}\ket{kl}_{AB}\right)$}
    \begin{subtable}[t]{0.47\textwidth}
        %\caption{For $i = 0$ to $i = 3$}
        \begin{tabular}{c|c|ccccccccc}
            \toprule
            $i$ & $k \setminus l$ & 0 & 1 & 2 & 3 & 4 & 5 & 6 & 7 & 8 \\
            \midrule
            & 0 & -3 & -3 & 0 & 0 & 0 & -3 & -3 & 0 & 0 \\
            0 & 1 & 0 & 0 & 0 & 4 & 4 & 0 & 0 & 4 & 0 \\
            & 2 & 0 & 0 & 0 & 0 & 0 & 0 & 0 & 0 & 0 \\
            \midrule
            & 0 & -1 & -1 & 0 & 0 & 0 & -1 & -1 & 0 & 0 \\
            1 & 1 & 0 & 0 & 1 & 1 & 0 & 0 & 0 & 1 & 1 \\
            & 2 & 0 & 0 & 0 & 0 & 0 & 0 & 0 & 0 & 0 \\
            \midrule
            & 0 & -3 & -3 & 4 & 0 & 0 & -3 & -3 & 0 & 0 \\
            2 & 1 & 4 & 0 & 0 & 0 & 0 & 0 & 0 & 0 & 0 \\
            & 2 & 0 & 4 & 0 & 0 & 0 & 0 & 0 & 0 & 0 \\
            \midrule
            & 0 & -3 & -3 & 0 & 4 & 0 & -3 & -3 & 0 & 0 \\
            3 & 1 & 0 & 0 & 0 & 0 & 0 & 0 & 4 & 0 & 0 \\
            & 2 & 4 & 0 & 0 & 0 & 0 & 0 & 0 & 0 & 0 \\
            \bottomrule
        \end{tabular}
    \end{subtable}
    %\hspace{0.02\textwidth}
    \begin{subtable}[t]{0.47\textwidth}
        %\caption{For $i = 4$ to $i = 7$}
        \begin{tabular}{c|c|ccccccccc}
            \toprule
            $i$ & $k \setminus l$ & 0 & 1 & 2 & 3 & 4 & 5 & 6 & 7 & 8 \\
            \midrule
            & 0 & -1 & -1 & 0 & 0 & 2 & -1 & -1 & 0 & 0 \\
            4 & 1 & 0 & 0 & 0 & 0 & 0 & 2 & 0 & 0 & 0 \\
            & 2 & 0 & 0 & 0 & 0 & 0 & 0 & 0 & 0 & 0 \\
            \midrule
            & 0 & -1 & -1 & 0 & 0 & 0 & -1 & -1 & 0 & 0 \\
            5 & 1 & 0 & 2 & 0 & 0 & 0 & 2 & 0 & 0 & 0 \\
            & 2 & 0 & 0 & 0 & 0 & 2 & 0 & 0 & 0 & 0 \\
            \midrule
            & 0 & -3 & -3 & 0 & 0 & 0 & -3 & -3 & 4 & 0 \\
            6 & 1 & 0 & 0 & 0 & 0 & 0 & 0 & 0 & 0 & 4 \\
            & 2 & 0 & 0 & 0 & 0 & 0 & 0 & 4 & 0 & 0 \\
            \midrule
            & 0 & -3 & -3 & 0 & 0 & 0 & -3 & -3 & 0 & 4 \\
            7 & 1 & 0 & 0 & 4 & 0 & 0 & 0 & 0 & 0 & 0 \\
            & 2 & 0 & 0 & 0 & 0 & 0 & 4 & 0 & 0 & 0 \\
            \bottomrule
        \end{tabular}
    \end{subtable}
    
\end{table}

Now, using the same method in Section-\ref{Algorithm genuinity} we evaluate the columnwise vectorization map by the $a$ coefficients from Table-\ref{genuinely_local_basis_table_II}. We calculate the rank to be $64$. This result proves the subspace spanned by the set $\Omega_{\ssty 3}^{\ssty3}(\mathcal{G}\up{1}{\ssty +}_{\ssty 000},\ket{\eta\down{1}{\ssty 0}}\up{-1}{\move{-1}\ssty (0)+})$ is entangled in $A|BC$ bipartition. The rest of the proof is heavily discussed in the main letter. Similarly, we get the same rank while considering other two bipartitions.

\section{Proof of Theorem 6}

As discussed, it is enough to prove the strong nonlocality of the set $\mathcal{U}_{\ssty 000}$. Using Theorem 5, it becomes too lucid to verify, and the verification process becomes more transparent. However, it can be observed that the order of the reduced feature matrix increases to the fourth multiplicity of the local dimension. Consequently, it becomes very challenging to evaluate the associated series of matrices and, additionally, to determine the dimension of the subspace spanned by them manually. So, we employed a computaional algorithm that produces the columnwise vectorization map of the reduced feature matrices in each bipartitions. The rank of the resulting matrix corresponds to the dimension of the subspace spanned by the reduced feature matrices defined for any set of quantum states.  For the states of $\mathcal{U}{\ssty 000}$, this rank is computed to be $80$ by processing the $a$-coefficients from (\ref{strong_nonlocal_basis}) across the bipartitions $AB|C$, $BC|A$, and $CA|B$. The subsequent part of the theorem is discussed in the main letter, building on this result.


\begin{thebibliography}{100}
\bibitem{NonlocalityIndistinguishability}It is important to note that when we refer to a set as local, we specifically mean that it is locally distinguishable and nonlocal if it is locally indistinguishable. Accordingly, nonlocality refers to local indistinguishability. This usage occurs frequently throughout the letter. It is crucial not to confuse these with the terms as used in the context of Bell nonlocality, where they carry entirely different meanings.
\end{thebibliography}
\end{document}